\documentstyle[preprint,prb,aps,epsfig]{revtex}
\tightenlines

\newcommand{\r} {{\mathbf r}}
\newcommand{\ns} {n_{\sigma}}
\newcommand{\nspair}{\{ n_\sigma \}}
\newcommand{\nr}{n({\mathbf r})}

\newcommand{\brhoxc}{\bar{\rho}_{xc}}
\newcommand{\brhox}{\bar{\rho}_{x}}
\newcommand{\brhoc}{\bar{\rho}_{c}}
\newcommand{\k} {{\mathbf{k}}}

\newcommand{\Exc} {E_{xc}}

\newcommand{\exc}{\epsilon_{xc}}
\newcommand{\ex}{\epsilon_{x}}

\newcommand{\si}{\sigma i}
\newcommand{\taus}{\tau_{\sigma}}

\newcommand{\Fxc}{F_{xc}}
\newcommand{\Fx}{F_{x}}
\newcommand{\Fc}{F_{c}}

\begin{document}
\draft

\title{Two-dimensional limit of exchange-correlation energy functional 
approximations in density functional theory}

\author{Yong-Hoon Kim,$^{1}$
 In-Ho Lee,$^{2,\ast}$
 Satyadev Nagaraja,$^{2}$
 Jean-Pierre Leburton,$^{2}$
 Randolph Q. Hood,$^{3,\dagger}$ and
 Richard M. Martin$^{1}$}
\address{$^1$ Department of Physics,
 University of Illinois at Urbana-Champaign, Urbana, Illinois 61801}
\address{$^2$ Beckman Institute for Advanced Science and Technology,
 University of Illinois at Urbana-Champaign, Urbana, Illinois 61801}
\address{$^3$ Cavendish Laboratory, Madingley Road, Cambridge CB3 0HE, United
Kingdom}

\date{September 14, 1999}

\maketitle

\begin{abstract}
We investigate the behavior of three-dimensional (3D) exchange-correlation
energy functional approximations of density functional theory in anisotropic
systems with two-dimensional (2D) character. Using two simple models,
quasi-2D electron gas and two-electron quantum dot, we show a {\it
fundamental limitation} of the local density approximation (LDA), and its
semi-local extensions, generalized gradient approximation (GGA) and meta-GGA
(MGGA), the most widely used forms of which are worse
than the LDA in the strong 2D limit. The origin of these shortcomings is in the
inability of the local (LDA) and semi-local (GGA/MGGA) approximations to
describe systems with 2D character in which the nature of the
exchange-correlation hole is very nonlocal. Nonlocal functionals provide an
alternative approach, and explicitly the average density approximation (ADA)
is shown to be remarkably accurate for the quasi-2D electron gas system. Our
study is not only relevant for understanding of the functionals but also
practical applications to semiconductor quantum structures and materials
such as graphite and metal surfaces.  We also comment on the implication of
our findings to the practical device simulations based on the (semi-)local
density functional method.
\end{abstract}


\narrowtext

\section{Introduction}
\label{sec:intro}

In the Kohn-Sham (KS) density functional theory (DFT)\cite{HK64,KS65},
significant efforts have been devoted to improve the local density
approximation (LDA).\cite{KS65} One approach, the generalized gradient
approximation (GGA)\cite{Lan81,Per85,Bec88,PW91,PBE96},
has been successively improved for the last two decades and now is
approaching chemical accuracy (atomization energy errors of order 1
kcal/mol = 0.0434 eV) with further refinements in the so-called
meta-GGA (MGGA).\cite{PKZB99} The (M)GGA is desirable in that it leads
to better physical quantities for various systems of interest, while
it is still computationally cheap due to its semi-local nature. It is
clear, however, that any local or semi-local approximation cannot
fully reproduce the behavior of the exact nonlocal
exchange-correlation energy functional, so one needs to be aware of
the limitations of these approximation schemes and the situations
where they can break down.

In this article, we discuss one situation where the local and semi-local
approximations of the exchange-correlation energy functional inherently
break down:
systems with two-dimensional (2D) characteristics, which
is relevant to DFT computations of semiconductor devices or other physical
systems with 2D character. The original motivation of the current work is
recent developments in semiconductor nanotechnology that have achieved
quantum dots, which offer enormous technological prospects and allow the
study of novel physical phenomena due to dimensionality and electronic
correlation effects.\cite{Kou97} Quantum dots can be achieved by confining a
2D electron gas with patterned gates. Semiconductor quantum
devices in general involve large ranges of electron densities and density
gradients, and the effect of electron-electron interactions can be
important, so they provide ideal test cases of the approximate functionals
commonly used in DFT. However, although DFT has been already extensively
applied to the study of these systems, \cite{Nag97,Lee98} the validity of
conventional approximation schemes in these systems has not been fully
addressed.  Hence, we investigate the robustness of various density-based
three-dimensional (3D) local and semi-local exchange and correlation energy
functional approximations, LDA, GGA, and MGGA, in the 2D-limit using the
idealized quasi-2D electron gas and quantum dot systems. We show that there
are inherent limitations resulting from the local or semi-local nature of the
exchange-correlation hole in these approximations. Especially, we point out
that
within the restricted form of the GGA
it is very difficult to incorporate the necessary
requirement for the 2D limit while at the same time maintaining desirable
features of present functionals. We contrast the limitation of these local
and semi-local approximations with the nonlocal average density approximation
(ADA), and explicitly show the improvement by employing the ADA for the
quasi-2D electron gas system.

The organization of the paper is as follows. In Sec. \ref{sec:review}, we
review the features of the LDA, GGA, MGGA, and ADA necessary for our later
discussions. In particular, we emphasize that the approximations in these
functionals are essentially approximations to the exchange-correlation hole.
In Sec. \ref{sec:limit}, we first establish the limitation of the local and
semi-local approximations by considering the nature of the approximations to
the exchange-correlation holes in the 2D limit, and contrast them with the
nonlocal approximation (Sec. \ref{subsec:2D_limit}). In Sec.
\ref{subsec:2degas} we explicitly show this in the 2D homogeneous electron
gas with finite thickness: compared with the exact exchange energy which is
finite in the 2D limit,  the 3D LDA, GGA, and MGGA exchange energies incorrectly diverge to
negative infinity. Especially, we point out that the direction of the GGA
and MGGA correction to the LDA should be opposite to that of the current
forms. This is contrasted with the nonlocal ADA approximation which has a
finite 2D-limit. In Sec. \ref{subsec:qdot}, we investigate an idealized
quantum dot system. By varying the confinement strength along one direction,
the system changes its character from 3D to 2D. We show that, while the LDA,
GGA, and MGGA give satisfactory descriptions of the isotropic limit, they
again fail in the 2D-limit.
Present (M)GGA's are better than the LDA in the isotropic 3D-limit, but they
are again worse than the LDA in the 2D-limit.  In addition, we comment on
the validity of 2D and 3D DFT calculations of quantum dots at the
experimentally realistic range of anisotropy. In Sec. \ref{subsec:other}, we
discuss density functional calculations of two physical systems with 2D
characters, jellium surface and the graphite.  We conclude this paper by
summarizing the current work in Sec. \ref{sec:conclusion}.

\section{exchange-correlation energy functionals}
\label{sec:review}

The exchange-correlation energy may be written as the interaction energy
between the electron density $n(\r) = \sum_{\sigma=\uparrow,\downarrow}
\ns(\r)$ and the coupling-constant integrated exchange-correlation hole
\cite{Lan75,Gun76}
$\brhoxc([\{ \ns \}];\r,\r')$:
\begin{equation}
\label{eq:Exc}
\Exc[\nspair] =
\frac{1}{2} \int d^3\r \ \int d^3\r' \
\frac{n(\r) \ \brhoxc([\{ \ns \}];\r,\r')}{|\r-\r'|},
\end{equation}
\begin{equation}
\label{eq:rhoxc}
\brhoxc([\{ \ns \}];\r,\r')
= n(\r') \int_0^1 d\lambda \ \bigl[ g^{\lambda}([\nspair];\r,\r')-1 \bigr]
\equiv n(\r') \bigl[ \bar{g}([\nspair];\r,\r')-1 \bigr],
\end{equation}
where $g^{\lambda}([\nspair];\r,\r')$ is the pair-correlation function. We
adopt atomic units throughout the paper with $\hbar = e = m_e = 1$. The
exact exchange and correlation hole have several important physical
conditions that should be also observed by approximations such as the
negativity of the exchange hole,\cite{Per85,PW91}
\begin{equation}
\label{eq:rhox-}
\brhox([\{ \ns \}];\r,\r') < 0,
\end{equation}
and sum rules of the exchange and correlation holes\cite{Lan75,Gun76,Gun79}
\begin{eqnarray}
\label{eq:rhoxc_SUM}
\int d^3\r' \brhox([\{ \ns \}];\r,\r') = -1, \nonumber\\
\int d^3\r' \brhoc([\{ \ns \}];\r,\r') = 0.
\end{eqnarray}

Various approximation schemes based on density-derived variables are
attempts to approximate the exchange-correlation energy functional
\begin{equation}
\Exc[\nspair] = \int d^3\r \; n(\r) \: \exc([\nspair];\r),
\end{equation}
or exchange-correlation energy density functional
\begin{equation}
\label{eq:exc}
\exc([\nspair];\r) = \frac{1}{2} \int d^3\r'
\frac{\brhoxc([\nspair];\r,\r')}{|\r-\r'|},
\end{equation}
in terms of a function of density and/or other density related variables. In
the standard LDA, the exchange-correlation energy density is replaced by
that of the homogeneous electron gas at each point $\r$
\begin{equation}
\exc([\nspair];\r) \approx \exc^{LDA}(\{ \ns(\r) \})
= \exc^{hom}(\{ \ns(\r) \}),
\end{equation}
which can be interpreted as an approximation to the exchange-correlation
hole \cite{Gun76,Gun79}
\begin{equation}
\label{eq:rhoxc_LDA}
\brhoxc^{LDA}(\{ \ns(\r) \};\r,\r')
 = n(\r) \ \bigl[\bar{g}^{hom}( \{ \ns(\r) \};|\r-\r'|)-1 \bigr].
\end{equation}
It is important to notice that the local replacement of density prefactor
$n(\r')$ by $n(\r)$ in Eq. (\ref{eq:rhoxc_LDA}) leads to the LDA
exchange-correlation hole being spherical and centered on the electron,
while the exact one is centered at the another point (such as at the nucleus
position in an atom or molecule) and very asymmetric. Thus we might expect
the LDA hole to be a reasonable approximation when the exact
exchange-correlation hole is close to the electron. Since only the spherical
average of the exchange-correlation hole
\begin{equation}
\label{eq:rhoxc_SA}
\brhoxc^{SA}(\r,R)=\frac{1}{4 \pi} \int_{\Omega} d\r' \ \brhoxc(\r,\r'), \qquad
\Omega: |\r-\r'|=R
\end{equation}
influences the exchange-correlation energy, \cite{Gun76,Gun79} the
spherically symmetric nature of the LDA hole does not necessarily represent
a poor approximation. In addition, it is known that the LDA is a
surprisingly robust approximation scheme,
which may be undestood from the fact that its
exchange-correlation
hole satisfies the hole conditions Eqs. (\ref{eq:rhox-}) and
(\ref{eq:rhoxc_SUM}). \cite{Gun76,Gun79}

In the following sections, we examine important features of other
approximations,
including the GGA, MGGA, and ADA, which are relevant for our discussions in
later sections.

\subsection{Generalized Gradient Approximation}
\label{subsec:GGA}

Although the idea of utilizing density gradient information as a way to
improve the LDA was proposed in the original papers of Kohn and
Sham\cite{HK64,KS65}, it is only in the last decade or so in which
successful GGA functionals have appeared. This suggests that some of the
correct physics of the exchange-correlation effects, which were missing in
the original naive gradient expansion approximation (GEA), have been
incorporated in recent developments of the GGA. The most important step in
the development of the GGA was the recognition by Perdew and coworkers that
the exchange-correlation hole in the GEA does not correspond to a physical
hole, nor does it satisfy the negativity condition of the exchange hole
[Eq.(\ref{eq:rhox-})] and normalization conditions of the exchange and
correlation holes [Eq.(\ref{eq:rhoxc_SUM})].\cite{Per85,PW91,Per96a}
Following their argument, the GGA can be understood as an approximation of
the exchange-correlation hole in which the spurious long-range part of the
second-order GEA exchange-correlation GEA hole has been cut-off in the real
space to satisfy the conditions of Eq. (\ref{eq:rhox-}) and Eq.
(\ref{eq:rhoxc_SUM}). One should note that the GGA is based on the
modification of the LDA exchange-correlation hole, so its hole is local and
it
tends to be an improvement over the LDA when the LDA is a good
first-order approximation. \cite{Per97_mail}

Different GGA approaches can be compared by writing the GGA exchange-correlation
energy density in terms of the reference LDA exchange energy density
multiplied by the factor $F_{xc}$:\cite{PW91,PBE96}
\begin{equation}
\label{eq:exc_GGA}
\exc([\nspair];\r)
\approx \exc^{GGA}(\{ \ns(\r),|\nabla \ns(\r)| \})
\equiv \ex^{LDA} (\nr) \ \Fxc^{GGA}( \{ \ns(\r),|\nabla \ns(\r)| \} ),
\end{equation}
where $\epsilon_x^{LDA}(\nr) = -3 (3 \pi^2 n(\r))^{1/3}/(4 \pi)$.
$\Fxc^{GGA}$, can be naturally divided into two parts, exchange $\Fx^{GGA}$
and correlation $\Fc^{GGA}$. For exchange, because of the spin-scaling
relation,
\begin{equation}
E_x[\nspair] = \frac{1}{2} E_x[2 n_{\uparrow}] + \frac{1}{2} E_x[2 n_{\downarrow}],
\end{equation}
we only need to consider the spin-unpolarized case $\Fx^{GGA}(n,|\nabla
n|)$, which is in terms of the dimensionless reduced density gradient
\begin{equation}
\label{eq:s}
s=\frac{|\nabla n|}{2 (3 \pi^2)^{1/3} n^{4/3}},
\end{equation}
is $\Fx^{GGA}(s)$. Numerous gradient approximations for the exchange have
been proposed, and in this work we consider the three most successful and
popular ones by Becke (B88),\cite{Bec88} Perdew and Wang (PW91),\cite{PW91}
and Perdew, Burke, and Enzerhof (PBE).\cite{PBE96} In Fig.
\ref{fig:Fxc_gga}, we compare the $\Fx^{GGA}$'s of these three
approximations. Most other $\Fx^{GGA}$'s fall between B88-GGA and
PBE-GGA\cite{Per96b}, so the qualitative results obtained by employing other
functionals can be interpolated from the behavior of the B88-GGA and
PBE-GGA.

As shown in Fig. \ref{fig:Fxc_gga}, one can divide the GGA into two regions,
(i) small $s$ ($0 \lesssim s \lesssim 3$) and (ii) large $s$ ($s \gtrsim 3$)
regions. In region (i), which is relevant for most physical applications,
different $\Fx^{GGA}$'s have nearly identical shapes, which explains why
different GGA's give similar improvement for many conventional systems with
small density gradient contributions. \cite{small_s} Most importantly,
$\Fx^{GGA} \geq 1$, so all the GGA's leads to exchange energy lower than the
LDA. Typically there are more rapidly varying density regions in atoms than
condensed system, so this will lead to the lowering of the exchange energy in
atoms more than molecules and solids.\cite{Zup97} This results in the
reduction of binding energy, correcting the LDA overbinding and improving
agreement with experiment, which is one of the most important characteristics
of present GGA's.

In region (ii), the different limiting behaviors of $\Fx^{GGA}$'s result
from choosing different physical conditions for $s \rightarrow \infty$. In
B88-GGA,  $\Fx^{B88-GGA}(s) \sim s/ln(s)$ was chosen to give the correct
exchange energy density ($\epsilon_x \rightarrow -1/2r$).\cite{Bec88} In
PW91-GGA, choosing $\Fx^{PW91-GGA}(s) \sim s^{-1/2}$ satisfies the
Lieb-Oxford bound and the non-uniform scaling condition.\cite{PW91} In
PBE-GGA, the non-uniform scaling condition was dropped in favor of a
simplified parameterization with $\Fx^{PBE-GGA}(s) \sim const$.\cite{PBE96}
The fact that different physical conditions lead to very different behaviors
of $\Fx^{GGA}$'s in region (ii) not only reflects the lack of knowledge of
the large density gradient regions but also an inherent difficulty of the
density gradient expansion in this region: even if one GGA form somehow
gives the correct result for a certain physical property while others fail,
it is not guaranteed that the form is superior for other properties in which
different physical conditions prevail.

Correlation is more difficult to include, but its contribution to the total
energy is typically much smaller than the exchange, especially for systems
with large density gradients. Hence, the main qualitative results of this
work, which is concerned with the strong anisotropic 2D-limit, are
determined at the exchange level. For correlations, we employed PW91- and
PBE-GGA which are almost identical and designed to be turned off for large
density gradient region as shown in Fig. \ref{fig:Fxc_gga}-(b) for the
PBE-GGA correlation part enhancement factor $\Fc^{PBE-GGA}$. The fact that
correlation decreases with increasing gradients can be qualitatively
understood in that systems with large gradients have strong confining
potentials that increase level spacings and reduce the effect of
correlations.

\subsection{Meta-Generalized Gradient Approximation}
\label{subsec:MGGA}

One natural extension of the GGA is to employ the next higher order gradient
expansion variables, the Laplacian of the density $\nabla^2 \ns(\r)$ and the
orbital kinetic energy density
\begin{equation}
\tau_{\sigma}(\r) = \frac{1}{2} \sum_{i=1}^{occp.}|\nabla \psi_{\si}(\r)|^2,
\end{equation}
in addition to the density $\ns(\r)$ and density gradient $\nabla \ns(\r)$.
Several forms of this so-called meta-GGA (MGGA) have been suggested, and
here we only consider the recent work of Perdew and his coworkers (PKZB)
based on the PBE-GGA.\cite{PKZB99} They removed the dependence on the
Laplacian of the density by introducing a new variable
\begin{equation}
\bar{q} = \frac{3 \tau}{2 (3\pi^2)^{2/3} n^{5/3}} - \frac{9}{20} -
\frac{s^2}{12},
\end{equation}
which reduces to the dimensionless Laplacian of the density $q = \nabla^2
n/(4 (3 \pi^2)^{2/3} n^{5/3})$ in the slowly varying limit but remains
finite at a nucleus where $q$ diverges. Hence we can write PKZB-MGGA as,
\begin{equation}
\label{eq:exc_MGGA}
\exc([\nspair];\r)
\approx
\exc^{MGGA}(\{\ns(\r),|\nabla \ns(\r)|,\taus \})
\equiv
\ex^{LDA} (\nr) \ \Fxc^{MGGA} ( \{ \ns(\r),|\nabla \ns(\r)|, \taus \} ),
\end{equation}
similar to Eq. (\ref{eq:exc_GGA}). The enhancement factor $\Fx^{PKZB-MGGA}$
for the exchange is shown in Fig. \ref{fig:Fx_mgga} as a function of $s$ and
$\bar{q}$. Unlike PBE-GGA, PKZB-MGGA exchange energy functional satisfies
both the correct gradient expansion and the linear response limit for the
exchange, and its correlation energy functional is self-interaction free
for a single electron.
However, the relevant feature of the $\Fx^{PKZB-MGGA}$ for the current
discussion is that it (or its exchange-correlation hole) is still semi-local
and $\Fx^{PKZB-MGGA}$ is always larger than or equal to 1 since it is based
on the PBE-GGA (see Fig. \ref{fig:Fx_mgga} along the $s$-axis). So the
qualitative feature of the PKZB-MGGA is similar to that of the PBE-GGA, and
its exchange energy is always lower than the LDA exchange energy.

\subsection{Average Density Approximation}
\label{subsec:ADA}

Two decades ago, Gunnarsson {\it et al.} criticized the earlier gradient
expansion approaches because of their failure to satisfy the sum rule [Eq.
(\ref{eq:rhoxc_SUM})], and proposed two completely nonlocal approximation
schemes, the average density approximation (ADA) and the weighted density
approximation (WDA).\cite{Gun79} In the ADA, which has been utilized in this
work, the exchange-correlation hole [Eq. (\ref{eq:rhoxc})] is approximated
by
\begin{equation}
\label{eq:rhoxc_ADA}
\brhoxc^{ADA}(\{ \bar{n}_{\sigma}(\r) \};\r,\r')
 = \bar{n}(\r) \; \bigl[\bar{g}^{hom}(\{ \bar{n}_{\sigma}(\r)
\};|\r-\r'|)-1\bigr],
\end{equation}
leading to
\begin{equation}
\exc([\{ \ns \}];\r) \approx \exc^{ADA}(\{ \bar{n}_{\sigma}(\r) \})
= \exc^{hom}(\{ \bar{n}_{\sigma}(\r) \}),
\end{equation}
where
\begin{equation}
\label{eq:n_ADA}
\bar{n}(\r) = \int d^3\r' \; w(\bar{n}(\r);\r-\r') \; n(\r'),
\end{equation}
is a non-local functional of the density. The important point is the
nonlocal nature of the ADA exchange-correlation hole whose extent depends
not only upon the density at the observation point but upon a weighted
average around $\r$. The weight function $w$ could be chosen in several
ways. Gunnarsson {\it et al.} originally proposed a form based on the
information of the linear response limit of the homogeneous electron gas. We
follow their suggestion and use the weight function $w$ tabulated in their
paper,\cite{Gun79} with the Eq. (\ref{eq:n_ADA}) evaluated by the method
based on fast Fourier transforms.\cite{Sin93,Hoo97}

\section{Inherent limitation of the local and semi-local approximations
 in the anisotropic 2D-limit}
\label{sec:limit}

\subsection{Basic issues}
\label{subsec:2D_limit}

We first outline the underlying physics by considering the behavior of the
exchange-correlation hole of 3D electrons in the 2D-limit. The 2D-limit of a
3D density can be written as
\begin{equation}
\label{eq:n3d_2d}
n(\r) \rightarrow n^{2D}(\r^{2D}) \; \delta(z).
\end{equation}
If we employ the exact exchange-correlation hole [Eq. (\ref{eq:rhoxc})], the
exchange-correlation energy density [Eq. (\ref{eq:exc})] in the 2D-limit is
\begin{equation}
\label{eq:exc2D}
\exc([\nspair];\r) \rightarrow
\frac{1}{2} \int d^2\r'^{2D} \,
 \frac{n^{2D}(\r'^{2D}) \;
 \bigl[ \bar{g}([\nspair];\r^{2D},\r'^{2D})-1 \bigr]}
 {|\r^{2D}-\r'^{2D}|}
\end{equation}
which is finite. Note that the Dirac delta function has been removed through
the integration along the z-direction. On the other hand, when we employ the
local LDA or the semi-local GGA/MGGA, density prefactor $n(\r')$ in the
exact exchange-correlation hole expression [Eq. (\ref{eq:rhoxc})] is
replaced by $n(\r)$. In these cases the Dirac delta function in Eq.
(\ref{eq:n3d_2d}) will not be removed through the integration in the
evaluation of the exchange-correlation energy density as in Eq.
(\ref{eq:exc2D}) which results in the divergence
\begin{equation}
\label{eq:exc2D_local}
\exc^{LDA,(M)GGA}(\{ \ns(\r) \}) \rightarrow -\infty.
\end{equation}
In conclusion, we can expect the incorrect divergence of the (semi-)local
approximations due to the approximation of the exchange-correlation hole as
being (semi-)local.

Now, we contrast this with a nonlocal approximation. Specifically, we employ
the ADA, which has been described in Sec. \ref{subsec:ADA} and will be used
in the next subsection. In the ADA, the prefactor $n(\r')$ is replaced by
$\bar{n}(\r)$ as in Eq. (\ref{eq:n_ADA}), which is finite in the 2D-limit:
\begin{equation}
\bar{n}(\r) \rightarrow
\bar{n}(\r^{2D}) =
\int d^2\r'^{2D} \; w\bigl(\bar{n}(\r^{2D});\r^{2D}-\r'^{2D}\bigr) \; n(\r'^{2D}).
\end{equation}
So, we expect the exchange-correlation hole and especially the
exchange-correlation energy density in the ADA will correctly have a finite
2D limiting value:
\begin{equation}
\exc^{ADA}(\{ \bar{n}_{\sigma}(\r) \}) \rightarrow
\frac{\bar{n}(\r^{2D})}{2} \;
\int d^2\r'^{2D} \,
\frac{\bigl[ \bar{g}^{hom}({\bar{n}_{\sigma}(\r^{2D})};|\r^{2D}-\r'^{2D}|)-1 \bigr]}
	{|\r^{2D}-\r'^{2D}|}.
\end{equation}

\subsection{Quasi-2D electron gas}
\label{subsec:2degas}

The 2D electron gas is experimentally realizable, for example, in the silicon
metal-oxide-semiconductor field-effect transistor (MOSFET) inversion layer
and in $GaAs/Al_xGa_{1-x}As$ heterostructures.\cite{And82}
Although the LDA in DFT formalism has been typically employed for the study
of many-body effect in these device systems since the 1970's,\cite{And82}
the limitations of the LDA has not been fully resolved.\cite{And82,Ste84} In
this section we investigate the accuracy of the LDA, GGA, and MGGA for the
quasi-2D homogeneous electron gas, which is an idealized model of a quantum
well.  For carrier with an isotropic effective mass $m^\ast$, such as
electrons in $GaAs/Al_xGa_{1-x}As$, interacting with
Coulomb interactions screened by a dielectric constant $\epsilon$,
the hamiltonian is the same as for electrons in free
space if one adopts scaled units of length
\begin{equation}
\label{eq:length}
\widetilde{\r}= \alpha \r; \: \alpha = \frac{m^\ast}{\epsilon},
\end{equation}
and energy
\begin{equation}
\label{eq:energy}
\widetilde{E} = \beta E; \: \beta = \frac{\epsilon^2}{m^\ast}.
\end{equation}
For $GaAs$, $m^\ast$ = 0.067 $m_e$ and $\epsilon$ = 13.2, hence the
effective unit of energy is 1 $\widetilde{Ha}$ = 10.46 meV and length is 1
$\widetilde{a}_0$ = 104.22 $\AA$. However, in the following, we drop
the tilde symbol for simplicity, unless explicitly stated otherwise.

A strict 2D electron gas can be characterized by one dimensionless
parameter, $r_s^{2D}=1/\sqrt{\pi n_A}$ or $k_F^{2D}= { 2\pi/ r_s^{2D}}$,
where $n_A$ is the areal electron number density, which ranges $10^{11} \sim
10^{13} \; cm^{-2}$ for typical $GaAs/Al_xGa_{1-x}As$ heterostructures.
However, since we are primarily interested in how various 3D DFT
exchange-correlation energy approximations perform in the 2D-limit, we
incorporate the finite thickness by including an envelope wavefunction
$\zeta_0(z)$, with the $z$-direction taken to be perpendicular to the 2D
homogeneous electron gas layer. Here, we assume that only the lowest single
2D subband is populated. Then, within the effective mass approximation, the
layer electrons can be characterized, by wavefunctions of the form
\begin{equation}
\label{eq:envelope_wf}
\psi(\r) = {1 \over \sqrt{A}} \; exp(i \k^{2D} \cdot \r^{2D})\; \zeta_0(z),
\end{equation}
where $\r^{2D}$ is the 2D radius vector and $\k^{2D}$ is the 2D wavevector.
It is normalized to area $A$, and $\zeta_0(z)$ is also assumed to be
normalized. We take the quantum well potential along the confinement
$z$-direction to be parabolic, with the envelope wavefunction $\zeta_0(z)$ of
the form
\begin{equation}
\label{eq:gaussian}
\zeta_0(z) = \bigl( \frac{b^2}{\pi} \bigr)^{1/4} \; e^{-b^2 z^2/2}.
\end{equation}
In Eq. (\ref{eq:gaussian}) $1/b$ characterizes the spatial extension of
wavefunctions along the $z-$direction, so we choose the dimensionless ratio
$b/k_F^{2D}$ as a measure of the finite thickness effect. Defining the
average thickness as $l_0 = \sqrt{2}/b$, which ranges from 20 to 100 $\AA$
in experiments, the physically relevant range of $b/k_F^{2D} = r_s^{2D}/l_0$
will be approximately between 1 and 5.

Assuming that the wavefunction has the form in Eq. (\ref{eq:gaussian}), Fig.
\ref{fig:2degas}-(a) and (b) show the comparison of the ratios of the
exchange and exchange-correlation energy per electron obtained from 3D exact
exchange (Hartree-Fock) method and various local/semi-local DFT
approximation schemes over the absolute value of the 2D exact exchange
energy. The quatntity plotted is the ratio of the energy to the absolute
value of the exchange energy in the 2D (large b) limit, $\epsilon_x^{exact,2D}
= -(4\sqrt{2})/(3\pi r_s^{2D})$. The finite thickness of the wavefunction
gives the correction $Y(b/k_F^{2D})$
\begin{equation}
\label{eq:2D_Ex}
\epsilon_x^{exact,3D} = - \frac{4\sqrt{2}}{3\pi r_s^{2D}} \;
Y\bigl(\frac{b}{k_F^{2D}}\bigr).
\end{equation}
The finite thickness correction $Y(b/k_F^{2D})$ in Eq. (\ref{eq:2D_Ex}) has
been calculated with the envelope-wavefunction Eq. (\ref{eq:envelope_wf}) in
a similar manner to Ref. \onlinecite{Ste74} where the Fang-Howard
envelope-wavefunction was used.\cite{And82} Physically, this finite
thickness correction makes the effective interaction softer than $1/r$
Coulomb interaction for distances small compared to the extension in the $z$
direction of the charge distribution, which leads to a significant correction to
the 2D exchange energy for
$b/k_F^{2D} \lesssim 10$.

The very weak confinement regime ($b/k_F^{2D} < 1$) in Fig.
\ref{fig:2degas}-(a) reveals that the LDA, GGA's, and MGGA exchange are very
close to the exact
exchange values. However, as we increase the confinement
strength they all go to the wrong limit: LDA, PBE-GGA, B88-GGA, and
PKZB-MGGA diverge to $-\infty$. Although PW91-GGA approaches a finite value
at large $b/k_F^{2D}$
(due to the fact that $\Fx^{PW91-GGA}(s) \sim s^{-1/2}$ for $s \rightarrow
\infty$),
the magnitude of the converged value is much too
large. Note that the LDA is better than the GGA or MGGA for the physically
relevant intermediate confinement strength regime, and the direction of
GGA/MGGA correction should be opposite to that of current forms, i.e., the
factor $F_x^{GGA}/F_x^{MGGA}$ should be less than 1 to reduce the error in
the LDA exchange. This feature is closely related with the nonuniform
scaling relation, and the PW91-GGA has this property built in, but only for
large-$s$ region.\cite{PW91} In order to describe quasi-2D quantum
nanostructures, one needs $F_x \leq 1$ for the small $s$ region, but this
will apparently worsen the (M)GGA for other systems such as spherical atoms.
This shows that it is difficult use the restricted (M)GGA form
to improve both 2D and conventional systems.

The total exchange-correlation energy for $r_s$ = 1.7 $\widetilde{a}_0$
($n_A$ = $10^{11} cm^{-2}$) is shown in Fig. \ref{fig:2degas}-(b), together
with the quasi-exact 2D-limit value obtained from the quantum Monte Carlo
calculations.\cite{Kwo93} The contribution of correlation energy is smaller
than the exchange energy in the physically relevant areal density regimes (5
$\sim$ 25 $\%$ for $n_A$ = $10^{13} \sim 10^{11}$ $cm^{-2}$), so the above
qualitative conclusions at the exchange level will not be changed with the
inclusion of the correlation energy. One noticeable difference between the
LDA and (M)GGA correlation energy functionals is that the magnitude of the
LDA correlation energy increases with increasing confinement strength, while
that of the (M)GGA decreases due the nature of their correlation form as
described in Sec. \ref{subsec:GGA}.

Now, we explicitly show the different behavior of the nonlocal approximation
which was expected in Sec. \ref{subsec:2D_limit}, by performing the 3D ADA
calculations on the quasi-2D electron gas model. The ratio of the ADA
exchange-correlation energy and the 2D exact exchange energy is shown in
Fig. \ref{fig:2degas}-(c) together with 2D LDA, and 3D exact exchange, LDA,
and PBE-GGA results. The ADA not only correctly reduces to a finite value
but also the limit value itself is surprisingly close to the exact 2D-limit.
This correct limiting behavior of the ADA clearly differentiate it from
local or semi-local approximations, and confirms our statements in Sec.
\ref{subsec:2D_limit}.
Recently, the ADA has been also shown to give improved descriptions of the
exchange-correlation energy density over the LDA for the conventional bulk
silicon system,\cite{Hoo97} but its applications are scarce in the
literature due to the difficulty of its implementation.

\subsection{Quantum Dot}
\label{subsec:qdot}

The second physical system with 2D character we discuss is a $GaAs$ quantum
dot. Because the orthogonal dot-growth direction confinement is usually much
stronger than the in-plane confinement, the electron distribution in a
quantum dot has a pancake-like spheroidal shape. For realistic potential
shapes of actual quantum dots, we refer the reader to Fig. 2 and 5 of Ref.
\onlinecite{Nag97}.

We take the simplest model of quantum dot systems with the external
potential an anisotropic harmonic oscillator potential, as
employed in our previous investigation,\cite{Lee98}
\begin{equation}
V_{ext}(\r) = \frac{1}{2} \omega^2 (x^2 + y^2) + \frac{1}{2} \omega_z^2 z^2.
\end{equation}
Here $z$ is the dot-growth direction, and we will examine $\omega_z \geq
\omega$. We consider only two interacting electrons in this potential. The
exact exchange energy for this two-electron system is equal to one half the
Hartree energy. In the isotropic potential limit $\omega = \omega_z$, this
model can be solved analytically for a discontinuous but infinite set of
oscillator frequencies,\cite{Tau93} and a comparative study of the exact KS,
LDA, and GGA schemes has been reported recently.\cite{Fil94} These results
were used as a check of our calculation method described below.

We performed self-consistent LDA, PBE-GGA, and exact exchange (EXX)
\cite{note:EXX} calculations for $\omega_z/\omega \leq 20$,  and PW91-GGA
and PKZB-MGGA energy have been evaluated by PBE-GGA density and wave
functions.
Technical details of our self-consistent EXX calculations based
on the finite-difference grid
scheme\cite{Lee98,YHK2} have been presented in Ref. \onlinecite{YHK4}.
For
$\omega_z/\omega > 20$, we used a simple variational EXX approach to
generate approximate solutioins: Taking the variational trial wavefunction
as
\begin{equation}
\psi(\r)= \bigl(\frac{\omega'}{\pi}\bigr)^{1/2} \; e^{-\omega' (x^2+y^2)/2} \;
 \bigl(\frac{\omega_z'}{\pi}\bigr)^{1/4} \; e^{-\omega_z' z^2/2}
\end{equation}
with two variational parameters $\omega'$ and $\omega_z'$ for a given
external potential characterized by $\omega$ and $\omega_z$, we minimize the
EXX total energy of the system, and use the variationally optimal
wavefunctions and the corresponding density to evaluate various 3D exchange
(and correlation) energy. The self-consistent and variational EXX exchange
energy obtained through this procedure shows the agreement of respectively
99.8 and 99.4 $\%$ with the exact KS value for the isotropic
case.\cite{Fil94} In addition to the 3D DFT calculations, we also performed
2D LDA calculations\cite{Kwo93} with the 2D density
\begin{equation}
n^{2D}(\r^{2D}) = 2 \; \bigl(\frac{\omega''}{\pi}\bigr) \;
 e^{-\omega'' (x^2+y^2)},
\end{equation}
that has been obtained in a 2D EXX variational minimization procedure as in
the large-confinement 3D case with a single variational parameter $\omega''$.

The exchange and exchange-correlation energies for $\omega$ = 2 meV = 0.1912
$\tilde{Ha}$ are shown in Fig. \ref{fig:qdot}-(a) and \ref{fig:qdot}-(b)
respectively. The EXX exchange values in Fig. \ref{fig:qdot}-(a) reveal that
the system approach the 2D limit at $\omega_z/\omega \sim 20 - 30$, and also
the 2D LDA exchange value is quite close to the EXX value in that limit. Now
we compare the 3D LDA, (M)GGA exchange energies with the EXX exchange
energy. First, at the isotropic limit ($\omega_z/\omega = 1$), we can see
that the (M)GGA improves the 3D LDA, since the magnitude of the exact
exchange energy is larger than that of the 3D LDA exchange energy at
$\omega_z/\omega = 1$. On the other hand, as we go to the more anisotropic
limit (larger $\omega_z/\omega$ regime),the 3D LDA and GGA exchange energies
become much larger in magnitude than the exact EXX values, and the GGA's
worsen the agreement with EXX, as in the 2D electron gas considered in the
previous subsection. The required directions of correction to the LDA value
have been indicated by upward/downward arrows, which shows that we need
corrections of of opposite signs at the isotropic and very anisotropic
limits;
however, the (M)GGA correction always has the same sign, which clearly shows
the inherent difficulty in constructing (M)GGA's that can satisfy all
limits.

The qualitative results at the exchange level are not changed with the
inclusion of the correlation energy as shown in Fig. \ref{fig:qdot}-(b),
except that, for small $\omega_z/\omega$, the differences between the 3D LDA
and (M)GGA exchange-correlation energies are smaller than the exchange case
alone, which shows that the 3D LDA is a good approximation with the exchange
and correlation together rather than separately.

Before closing this section, we comment on realistic experimental situations
for quantum dots. Using a value of $\omega_z = 20 - 50$ meV, estimated by
the vertical extent of quantum dots (approximately $100 \AA$), one finds the
realistic range of $\omega_z/\omega$ is about $10-25$, which is not the
extreme 2D-limit where we observed the breakdown of 3D local/semilocal DFT
approximations. Hence 3D LDA or (M)GGA quantum dot simulation results should
be reliable for such experimentally realistic problems. Further Fig.
\ref{fig:qdot}-(b) suggests that 3D and 2D functionals are about equally
applicable in this range although 3D functionals are definitely better for
$\omega_z/\omega \lesssim 10$.

In addition, by comparing our 3D and 2D EXX models, we can study the effect
of employing strict-2D quantum dot simulations as have been adopted in many
theoretical studies.\cite{Kou97} In Fig. \ref{fig:2Dvs3D}, we compare the
lateral components of 3D total kinetic energy $\langle K_{||} \rangle$,
total external potential energy $\langle V_{ext} \rangle$, and Hartree $+$
exchange energy $\langle V_{ee} \rangle$, with the corresponding 2D values.
Physically, the electron-electron interaction makes the electrons more
extended than the noninteracting counterparts, which decreases the kinetic
energy and increases the potential energy. On the other hand, the finite
thickness of the 2D electron layer has the effect of softening the Coulomb
interaction as discussed in Sec. \ref{subsec:2degas}, which results in the
increase of the kinetic energy and decrease of the potential energy than the
2D limiting values as shown in Fig. \ref{fig:2Dvs3D}. This effect is
pronounced up to $\omega_z/\omega \sim 20$, and we expect this feature will be
missing in the strict-2D calculations. One more noticeable point is that the
anisotropy of the potential induces bigger change in the Hartree $+$
exchange energy (+1.20 meV) than those in the kinetic energy (-0.38 meV) and
potential energy (+0.80 meV) from the values of the isotropic limit.

\subsection{ Other systems }
\label{subsec:other}

In this subsection, we consider two physical systems with 2D character to
which our study might have relevance.  First example is jellium surfaces,
which have recently drawn renewed interest due to the discrepancies between
results from different schemes.\cite{Kro86,Aci96,Pir98} Although we believe
quatum Monte Carlo calculation results\cite{Aci96} are the most accurate, it
is not our intention here to address this question. Rather, the relevant
point we would like to emphasize is that the conclusions of the present
study are consistent with those of Rasolt and Geldart \cite{Ras75} on the
gradient corrections in the jellium surface: to support their first gradient
correction coefficient having a different sign from that found in the atomic
systems, they emphasized the difference between the localized and extended
system and argued that their form is the proper one to use especially for
jelliumm surfaces.\cite{Ras86} Actually, they further brought caution on
forcing a ``universal'' gradient approximation form where none exists, which
is a warning we feel has been largely ignored. We believe our study supports
the argument of Rasolt and Geldart, and especially our second model system,
quantum dot, is a dramatic example showing the nonexistence of a universal
GGA form.

The second system we examine is graphite. Graphite and intercalated graphite
constitute another large family of materials with quasi-2D electronic
properties. Interestingly, in our previous invetigations, we observed that
the LDA gives a better description of the energy difference between the
diamond and hexagonal graphite structures of C than the PW91- or PBE-GGA
both at the theoretical and experimental lattice constants.\cite{Lee97}
Since the LDA and GGA descriptions of the diamond C are both satisfactory,
we can conclude that the GGA descriptions of the graphite is the origin of
the problem.  To trace the specific source of the error, we decomposed the
self-consistent total energies at the experimental lattice constants from
LDA, PW91-GGA, and PBE-GGA calculations into kinetic, potential, and
exchange-correlation energies. The kinetic + potential energy differences
between the diamond and graphite structures from the LDA/PBE-GGA
calculations were -820/-865 meV, while the exchange-correlation energy
differences between the two structures were 231/386 meV.  The fact that the
magnitude of the kinetic + potential energy is more than two times bigger
than the exchange-correlation energy, while the energy difference in the
exchange-correlation part is three times bigger than the kinetic + potential
part indicates that the description of the graphite in the PBE-GGA
functional is the main source of the deficiencies. This is closely related
with our findings in the present work, that the GGA gives poorer
descriptions of 2D systems than the LDA. In hexagonal graphite, there are
eight electrons per unit cell, and two of them in the $\pi$-like state can
be considered as 2D electrons, which corresponds to $r_s$ = 3.54
$\widetilde{a}_0$ or $k_F^{2D}$ = 0.41 $\widetilde{a}_0^{-1}$. The
valence-electron charge density distribution is known
experimentally,\cite{Che77} so one can estimate the thickness of the
electron layer $l_0$ as $\approx$ 2.33 $\widetilde{a}_0$ or $b =
\sqrt{2}/l_0 = 0.61 \; \widetilde{a}_0^{-1}$. This correspond to $b/k_F^{2D}
= 1.5$ in Fig. 3, suggesting that the GGA's poorer description of the energy
difference between the hexagonal graphite and diamond can be explained by
our findings in this work.

\section{ SUMMARY AND DISCUSSIONS }
\label{sec:conclusion}

The purpose of this work was to show the fundamental limitation of the 3D
local/semi-local exchange-correlation energy functional approximations of
DFT by considering systems with 2D characteristics. We traced the source of
the failure of the LDA, GGA, and MGGA in 2D systems to the (semi-)local
nature of their approximate exchange-correlation holes. These
local/semi-local approximations have been contrasted with a nonlocal
approximation such as the ADA, and the difference has been explicitly
demonstrated in our first model, quasi-2D electron gas, in which the ADA
reproduced the correct limiting behavior. The 2D-limit can be considered as
a constraint on approximate functionals. This condition is not built in most
of the (M)GGA's, and we emphasized that its form is inherently too
restricted to incorporate this requirement while keeping the necessary
property to improve the LDA for other conventional systems. Our second
example, an anisotropic quantum dot in which we need different signs of the
(M)GGA corrections to the LDA at 3D- and 2D-limits, explicitly shows the
danger of expecting a universal gradient approximation form as pointed out
by Rasolt and Geldart. \cite{Ras75} Two other realistic systems, jellium
surface and graphite have been discussed as relevant examples.  For
practical device simulations, however, we pointed out that the LDA and
(M)GGA results should be qualitatively correct, as long as experimentally
realistic situations are considered.

{\it Note} After completion of the present work, we were informed that
Pollack and Perdew have also studied the quasi-2D electron gas results using
a different quantum well model, finding results in agreement with ours
presented in Sec.\ref{subsec:2degas}. \cite{Pol00} They have employed a
scaling argument more extensively, and also made a connection to the liquid
drop model.

\acknowledgments
We thank Prof. K. Burke for valuable discussion and suggestion to
refer to Ref. \onlinecite{Zup97}, Prof. J. Perdew for helpful
discussion and sharing his works prior to publications, and V. Rao and
J. Shamaway for suggestions and performing some test calculations.
This work was supported by the National Science Foundation under Grant
No. DMR94-22496, DMR98-0273.

%

$^\ast$ Present address: School of Physics, Korea Institute for Advanced Study,
Cheongryangri-dong, Dongdaemun-gu, Seoul 130-012, Korea

$^\dagger$ Present address: Lawrence Livermore National Laboratory, Livermore,
CA 94550.


\newpage

\begin{figure}
\caption{The enhancement factor over the LDA exchange energy density for (a)
B88-, PW91-, and PBE-GGA exchange $F_x^{GGA}$ and (b) PBE-GGA correlation
$F_c^{PBE-GGA}$ in terms of the dimensionless reduced density gradient
$s={|\nabla n|}/(2 k_F n)$.}
 \label{fig:Fxc_gga}
\end{figure}

\begin{figure}
\caption{The exchange part of PKZB-MGGA enhancement factor over the LDA
exchange energy density $F_x^{PKZB-MGGA}$ in terms of the dimensionless
reduced density gradient $s={|\nabla n|}/(2 k_F n)$ and $\bar{q}$ defined in
the text.}
 \label{fig:Fx_mgga}
\end{figure}

\begin{figure}
\caption{ The ratio of 3D exact, LDA, GGA's and MGGA (a) exchange energy and
(b) exchange-correlation energy per electron over the absolute value of the
2D exact exchange energy for the quasi-2D jellium model versus $b/k_F^{2D}$.
(c) The ratio of the 3D ADA exchange-correlation energy per electron over
the absolute value of the 2D exact exchange energy for the same system. This
ratio is independent of the areal electron number density at the
exchange-only level (a), and $r_s$ = 1.7 $\widetilde{a}_0$ ($n_A$ =
$10^{11}$ $cm^{-2}$) case has been shown in (b) and (c). The experimentally
realistic range of $b/k_F^{2D}$ is about $1-5$. For (c), the exact exchange,
LDA and PBE-GGA exchange-correlation energies given in (b) are reproduced
for comparison. Note that the ADA reduces very closely to the exact 2D
exchange-correlation limit.}
 \label{fig:2degas}
\end{figure}

\begin{figure}
\caption{(a) Exchange energy and (b) exchange-correlation energy of 2
electrons confined in a quantum dot modeled by an anisotropic harmonic
oscillator potential as functions of $\omega_z/\omega$. For a fixed
lateral-direction external potential $\omega$ = 2 meV, the dot-growth
direction potential $\omega_z$ has been varied from 2 to 2000 meV. The
realistic value of $\omega_z$ is about $20-50$ meV, or $\omega_z/\omega
\approx 10-25$.} The upward/downward arrows in (a) indicate the direction of
correction to the LDA exchange energy.  Note that (M)GGA correction is
always downward, hence improves the LDA at the isotropic limit, but worsens
at the anisotropic limit.
 \label{fig:qdot}
\end{figure}

\begin{figure}
\caption{The lateral components of 3D total kinetic energy $\langle K_{||}
\rangle$, total external potential energy $\langle V_{ext} \rangle$, and
Hartree $+$ exchange energy $\langle V_{ee} \rangle$, and the corresponding
2D values $\langle K^{2D} \rangle$, $\langle V_{ext}^{2D} \rangle$, and
$\langle V_{ee}^{||} \rangle$ obtained from variational EXX calculations.
For a fixed lateral-direction confinement potential $\omega$ = 2 meV, t.he
dot-growth direction potential has been varied from 2 to 200 meV.}
 \label{fig:2Dvs3D}
\end{figure}

\newpage

  \begin{minipage}[H]{0.70\linewidth}
  \vspace{0.5cm}
  \centering\epsfig{file=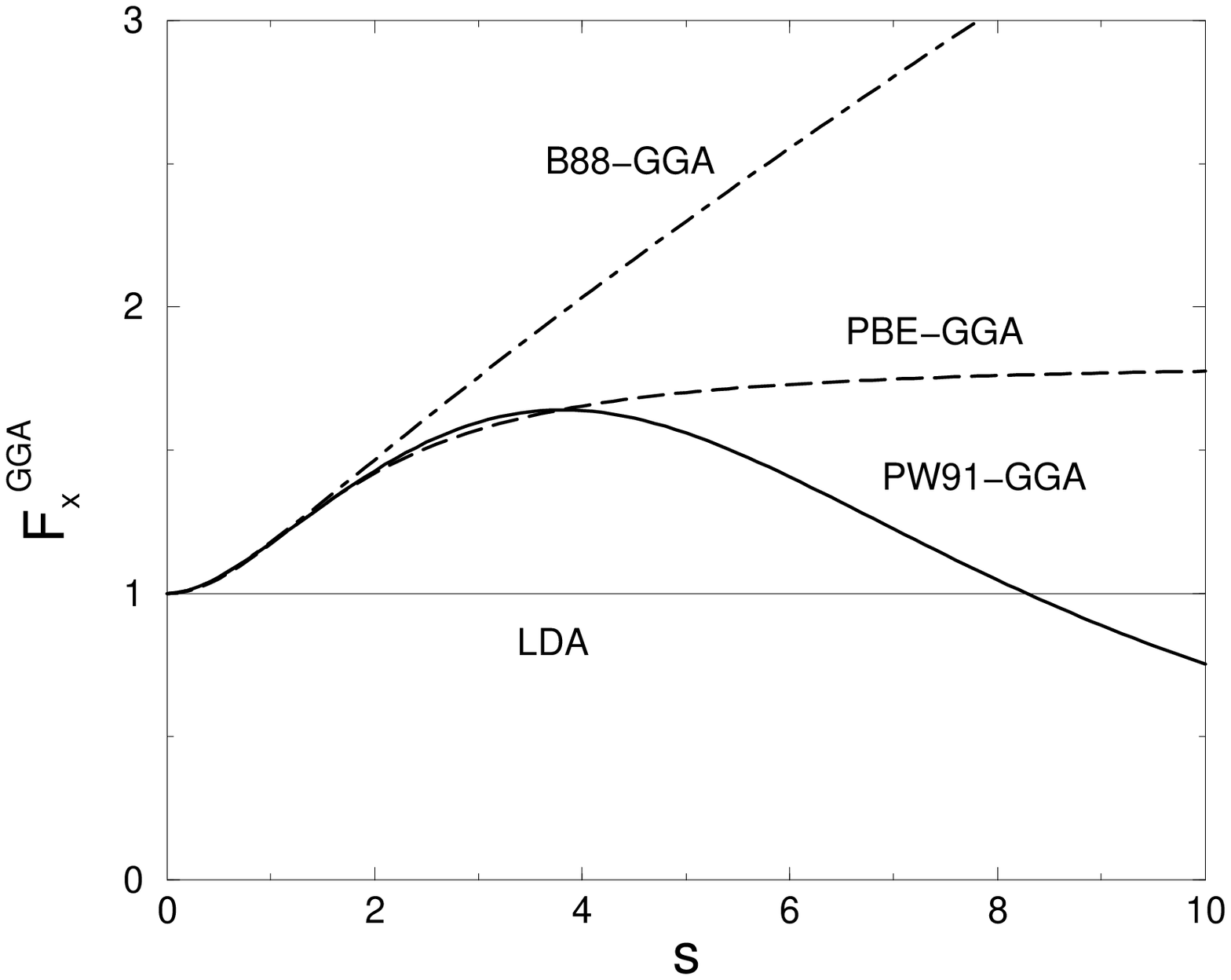,width=\linewidth}
  \end{minipage}  \hspace{1.2cm}
   Fig. 1 (a)

  \begin{minipage}[H]{0.70\linewidth}
  \vspace{0.5cm}
  \centering\epsfig{file=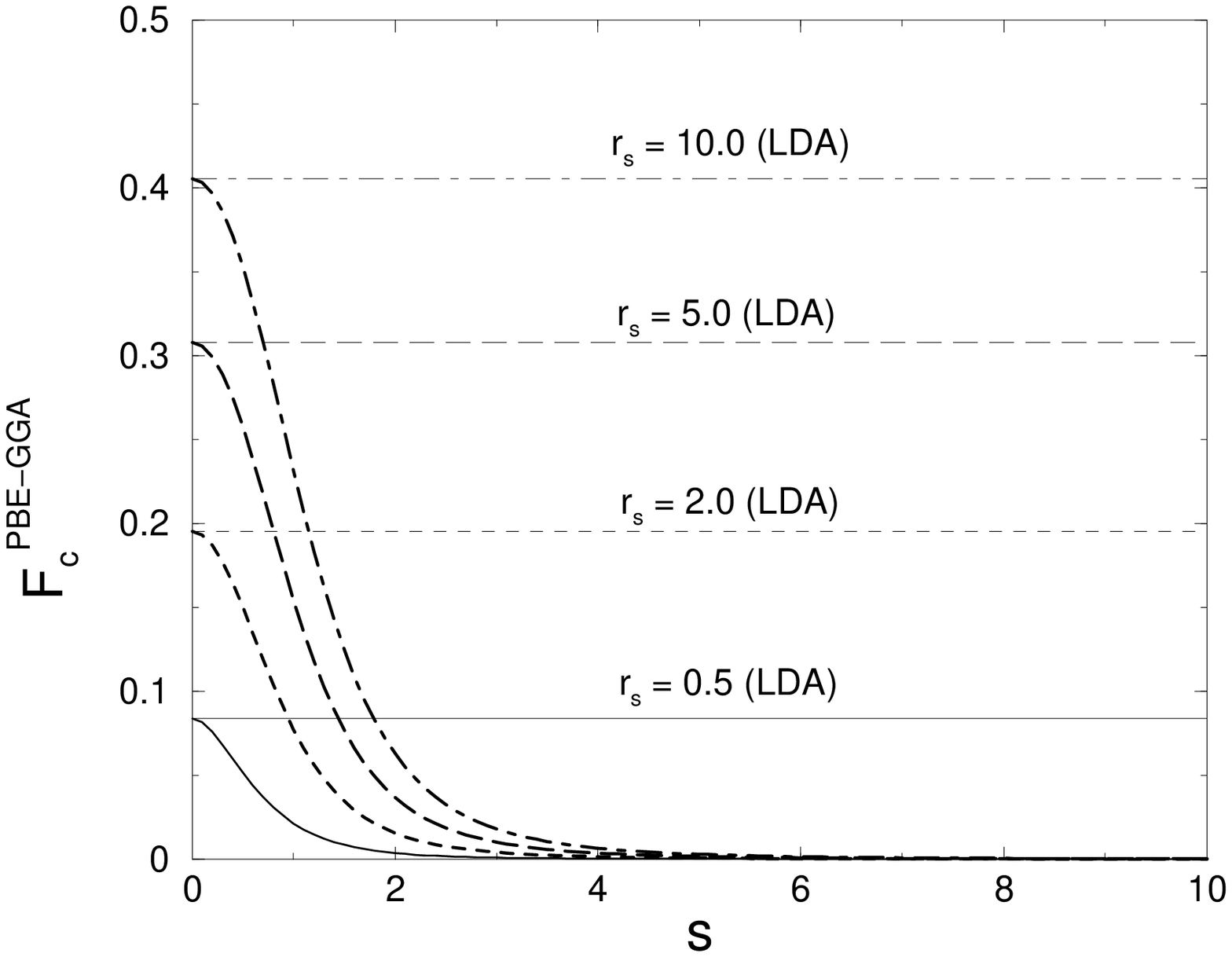,width=\linewidth}
  \end{minipage}  \hspace{1.2cm}
   Fig. 1 (b)

  \begin{minipage}[H]{0.70\linewidth}
  \vspace{0.5cm}
  \centering\epsfig{file=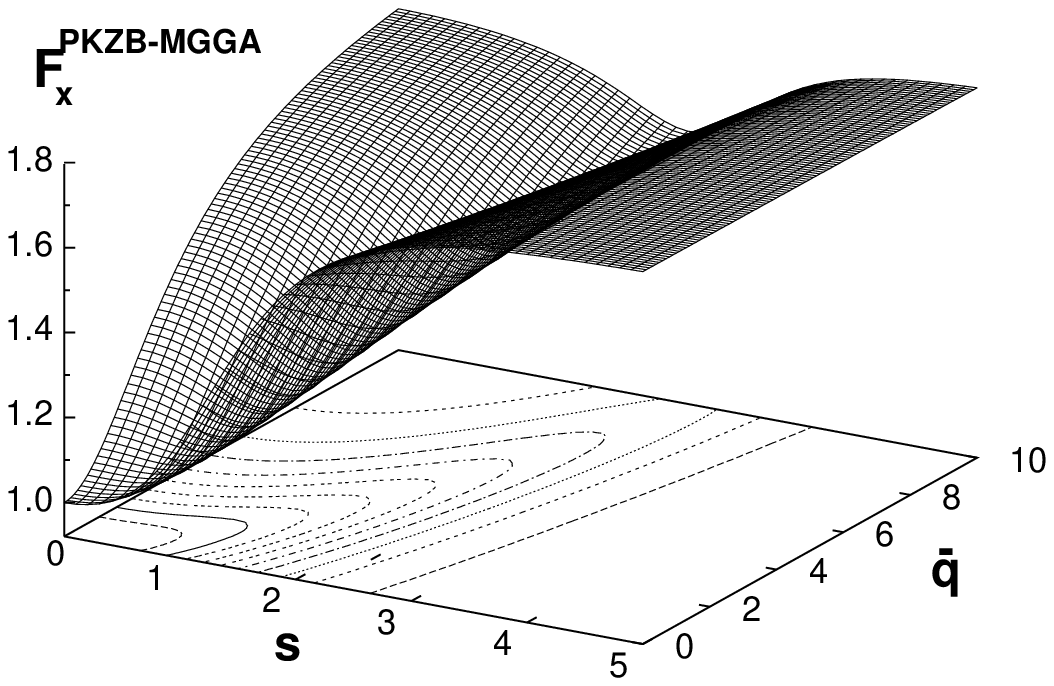,width=\linewidth}
  \end{minipage}  \hspace{1.2cm}
   Fig. 2

  \begin{minipage}[H]{0.70\linewidth}
  \vspace{0.5cm}
  \centering\epsfig{file=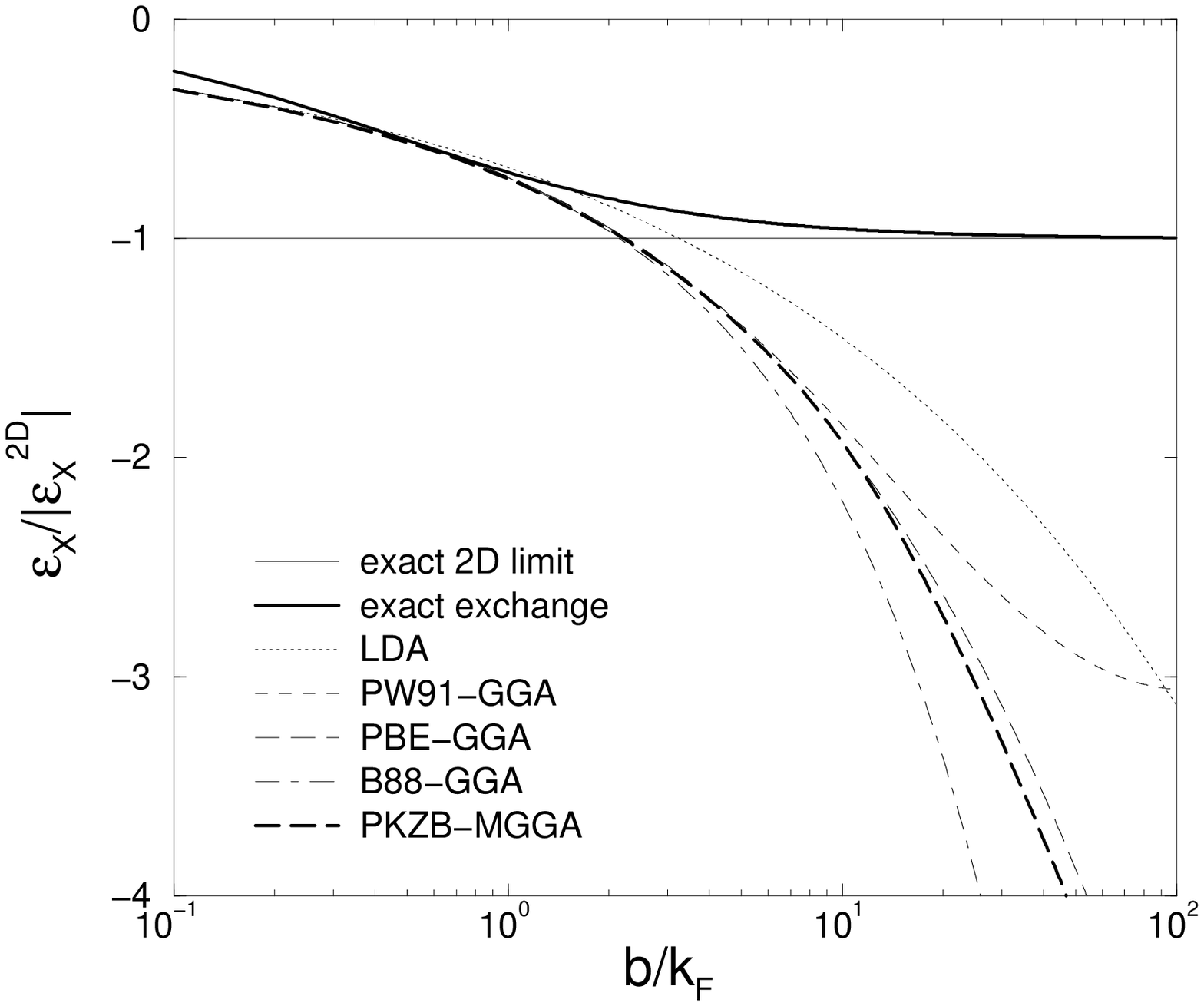,width=\linewidth}
  \end{minipage}  \hspace{1.2cm}
   Fig. 3 (a)

  \begin{minipage}[H]{0.70\linewidth}
  \vspace{0.5cm}
  \centering\epsfig{file=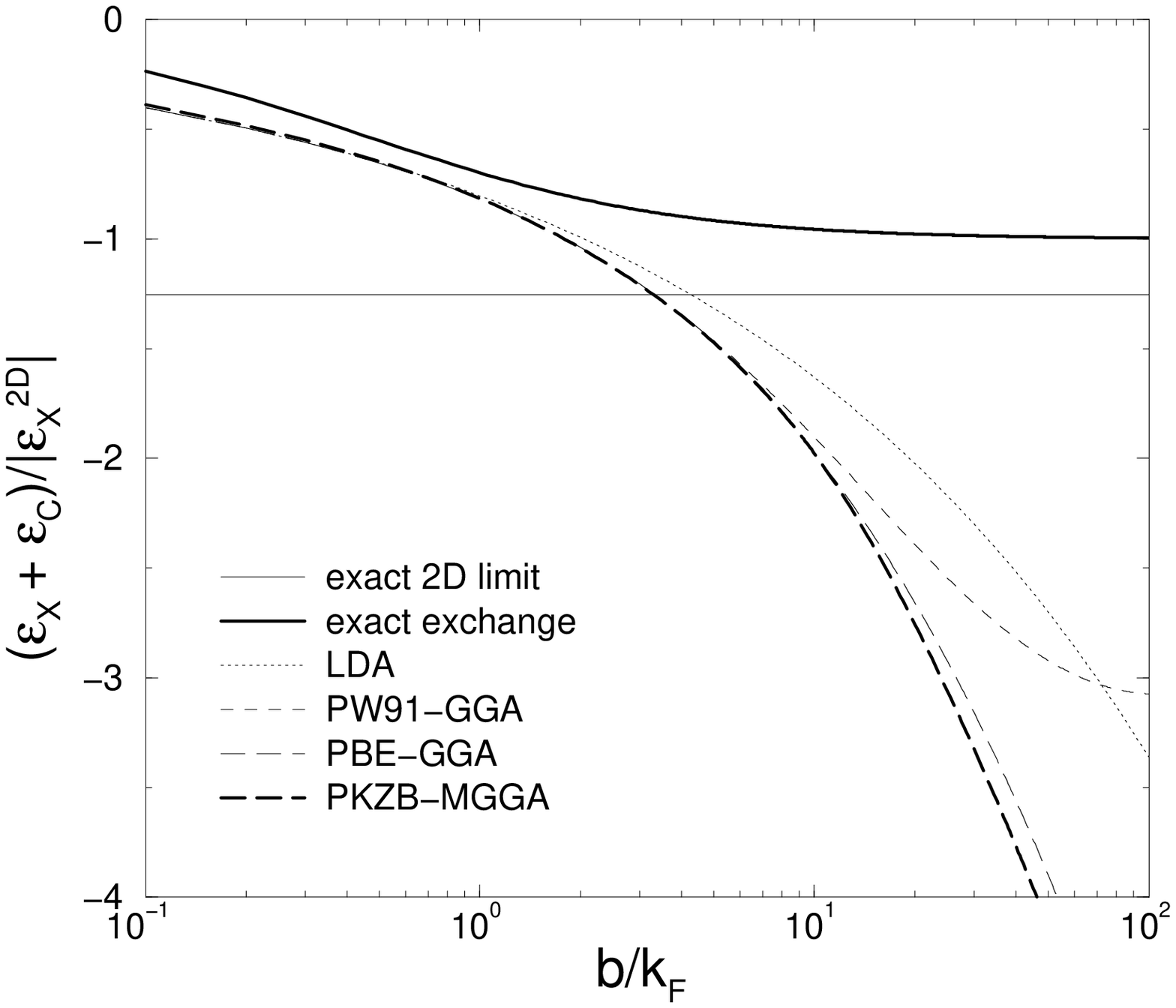,width=\linewidth}
  \end{minipage}  \hspace{1.2cm}
   Fig. 3 (b)

  \begin{minipage}[H]{0.70\linewidth}
  \vspace{0.5cm}
  \centering\epsfig{file=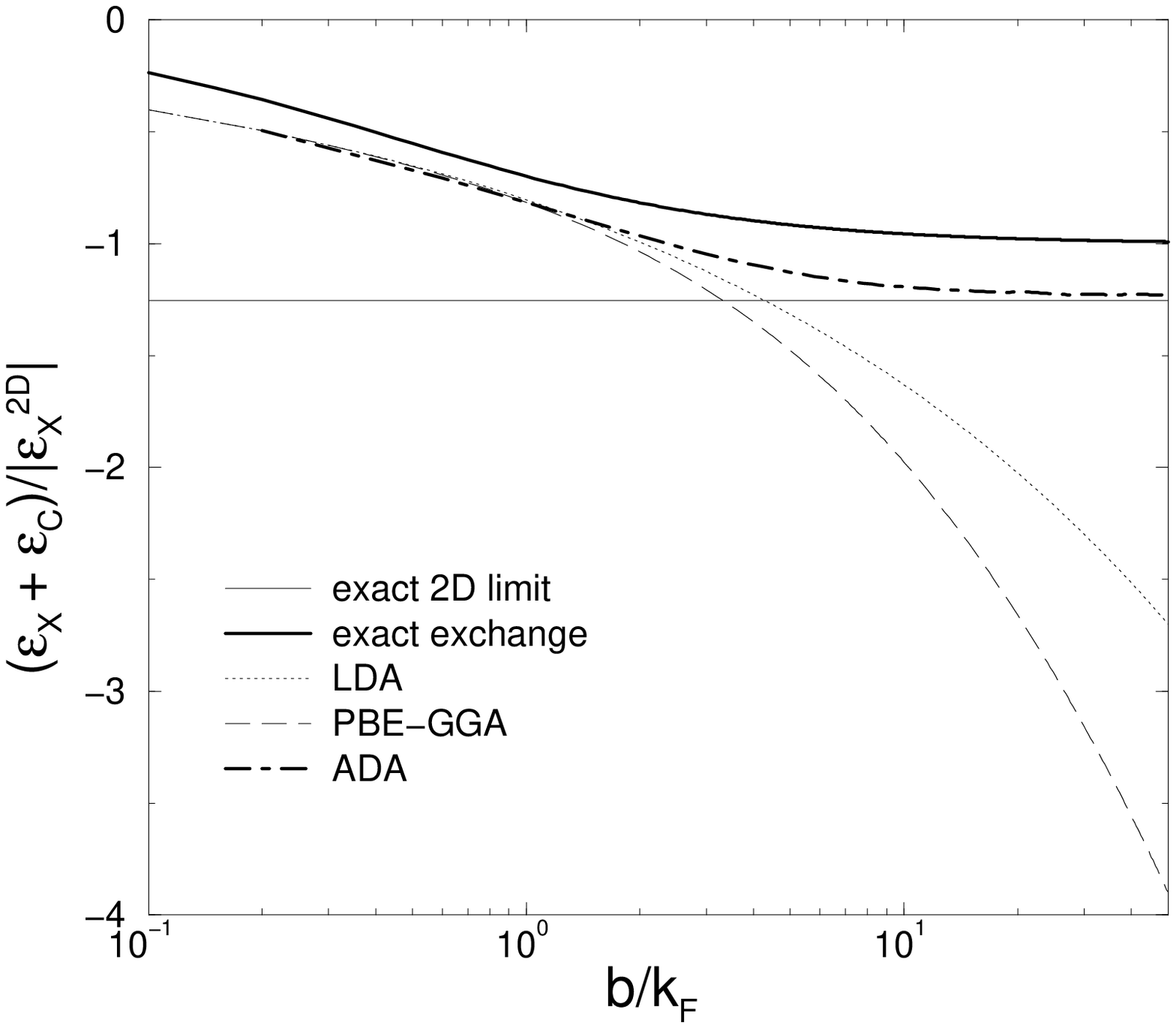,width=\linewidth}
  \end{minipage}  \hspace{1.2cm}
   Fig. 3 (c)

  \begin{minipage}[H]{0.70\linewidth}
  \vspace{0.5cm}
  \centering\epsfig{file=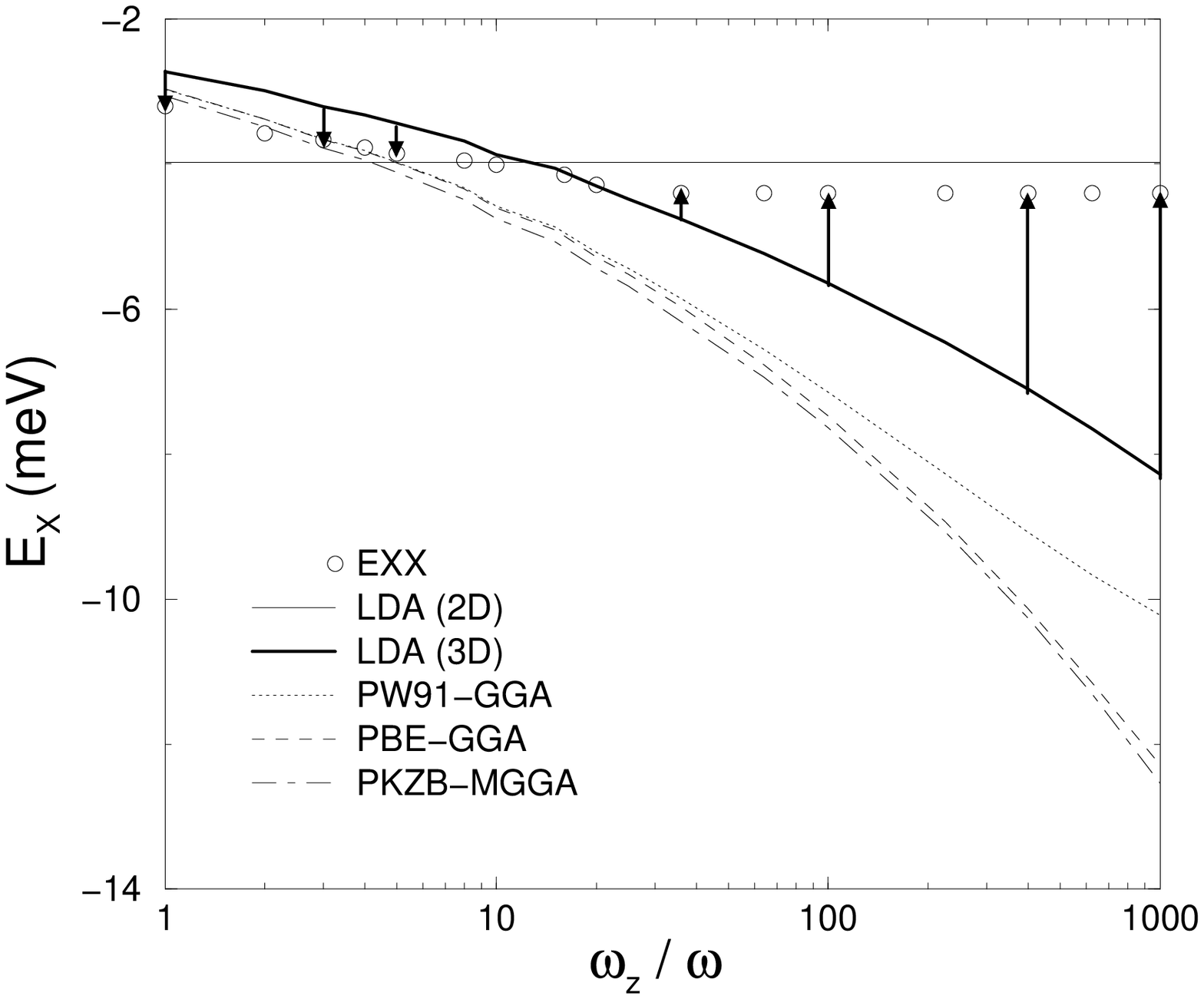,width=\linewidth}
  \end{minipage}  \hspace{1.2cm}
   Fig. 4 (a)

  \begin{minipage}[H]{0.70\linewidth}
  \vspace{0.5cm}
  \centering\epsfig{file=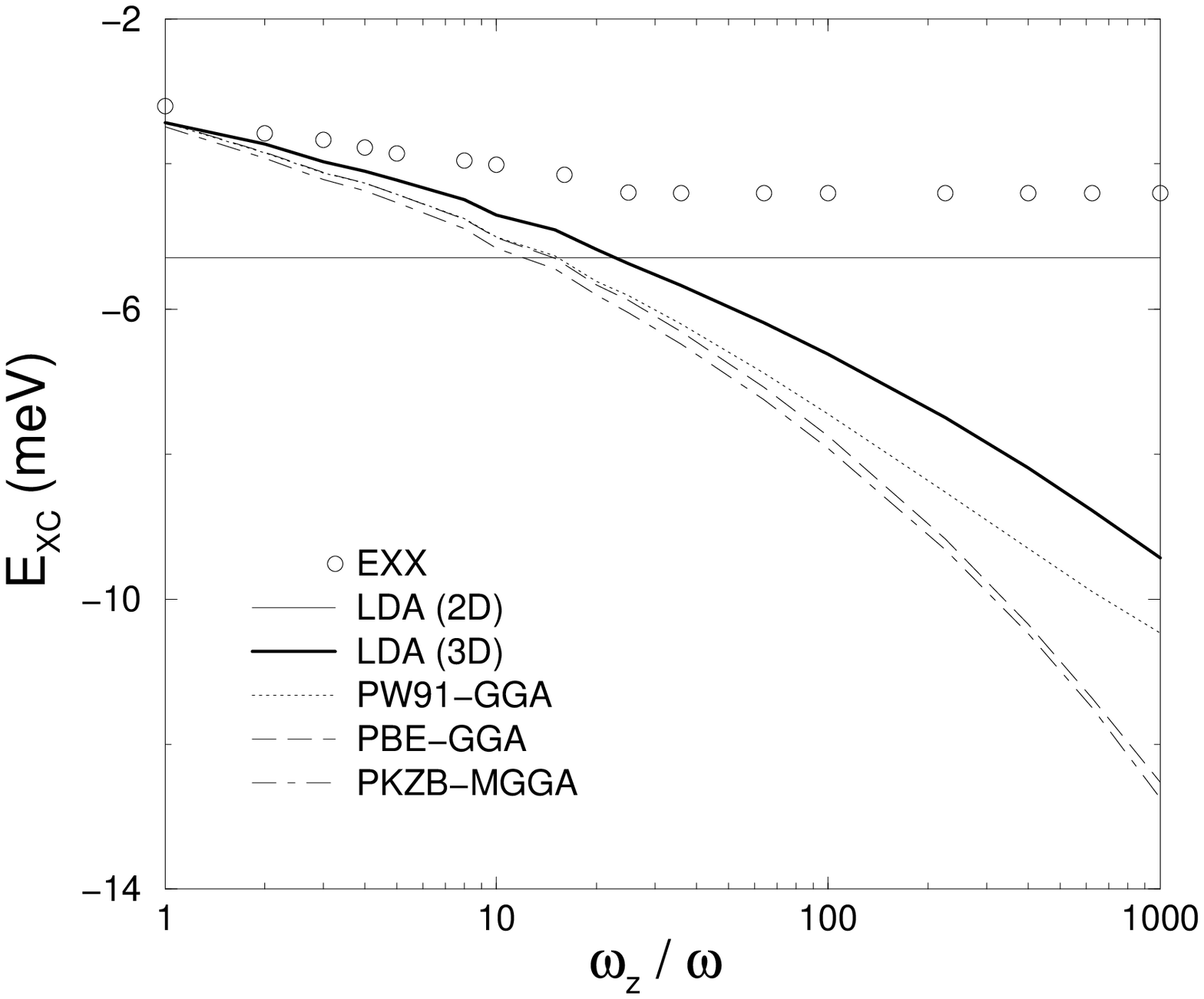,width=\linewidth}
  \end{minipage}  \hspace{1.2cm}
   Fig. 4 (b)

  \begin{minipage}[H]{0.70\linewidth}
  \vspace{0.5cm}
  \centering\epsfig{file=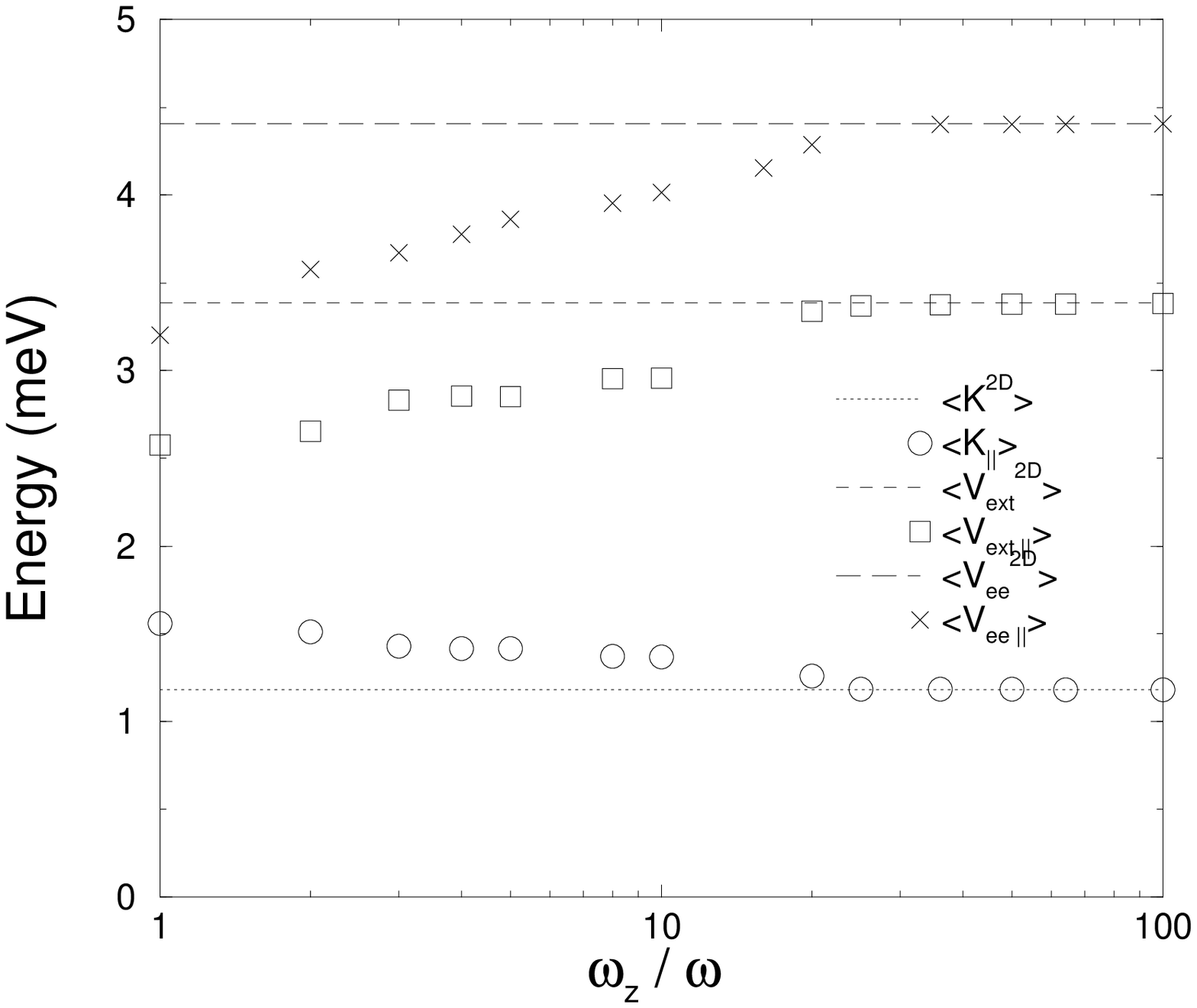,width=\linewidth}
  \end{minipage}  \hspace{1.2cm}
   Fig. 5


\begin{references}

\bibitem{HK64}
P. Hohenberg and W. Kohn,
Phys. Rev. {\bf 136}, B864 (1964).

\bibitem{KS65}
W. Kohn and L. J. Sham,
Phys. Rev. {\bf 140}, A1133 (1965).

\bibitem{Lan81}
D. C. Langreth and M. J. Mehl,
\prl {\bf 47}, 446 (1981);
\prb {\bf 28}, 1809 (1983).

\bibitem{Per85}
J. P. Perdew,
\prl {\bf 55}, 1665 (1985);
J. P. Perdew and Y. Wang,
\prb {\bf 33}, 8800 (1986).

\bibitem{Bec88}
A. D. Becke,
\pra {\bf 38}, 3098 (1988).

\bibitem{PW91}
K. Burke, J. P. Perdew, and Y. Wang,
in {\it Electronic Density Functional Theory: Recent Progress and New
Directions},
edited by J. F. Dobson, G. Vignale, and M. P. Das
(Plenum, NY, 1998).

\bibitem{PBE96}
J. P. Perdew, K. Burke, and M. Ernzerhof,
\prl {\bf 77}, 3865 (1996).

\bibitem{PKZB99}
J. P. Perdew, S. Kurth, A. Zupan, and P. Blaha,
\prl {\bf 82}, 2544 (1999).

\bibitem{Kou97}
L. P. Kouwenhoven, C. M. Marcus, P. L. McEuen, S. Tarucha, R. M. Westervelt, and
N. S. Wingreen,
{\it Mesoscopic Electron Transport}, edited by L. L. Sohn, L. P. Kouwenhoven,
and G. Sch\"on
(Kluwer Academic, Dordrecht, 1997).

\bibitem{Nag97}
S. Nagaraja, P. Matagne, V.-Y. Thean, and J.-P. Leburton, Y.-H. Kim, and R. M.
Martin,
\prb {\bf 56}, 15752 (1997).

\bibitem{Lee98}
I.-H. Lee, V. Rao, R. M. Martin, and J.-P. Leburton,
\prb {\bf 57}, 9035 (1998).

\bibitem{Lan75}
D. C. Langreth and J. P. Perdew,
Solid State Commun. {\bf 17}, 1425 (1975).

\bibitem{Gun76}
O. Gunnarsson and B. I. Lundqvist,
\prb {\bf 13}, 4274 (1976).

\bibitem{Gun79}
O. Gunnarsson, M. Jonson, B. I. Lundqvist,
\prb {\bf 20}, 3136 (1979).

\bibitem{Per96a}
J. P. Perdew, K. Burke, and Y. Wang,
\prb {\bf 54}, 16533 (1996).

\bibitem{Per97_mail}
J. P. Perdew, private communications.

\bibitem{Per96b}
J. P. Perdew and K. Burke,
Int. J. Quantum Chem. {\bf 57}, 309 (1996).

\bibitem{small_s}
While these three GGA functionals have similar shapes in region (i)
the detailed behaviors are different as $s \rightarrow 0$.
In B88-GGA,\cite{Bec88} parameters were chosen empirically resulting in
$\Fx^{B88-GGA}(s) \rightarrow 1 + 0.235 s^2$,
while in PW91-GGA,\cite{PW91} $s \rightarrow 0$ limit was chosen to reproduce
the
correct gradient expansion for small $s$: $\Fx^{PW91-GGA}(s) \rightarrow 1 +
0.123 s^2$.
In PBE-GGA,\cite{PBE96} the correct descriptions of the linear response of
a uniform electron gas was incorporated in preference to the correct
gradient expansion condition resulting in $\Fx^{PBE-GGA}(s) \rightarrow 1 +
0.220 s^2$.

\bibitem{Zup97}
A. Zupan, K. Burke, M. Ernzerhof, and J. P. Perdew,
\jcp {\bf 106}, 10184 (1997).

\bibitem{Sin93}
D. J. Singh, 
\prb {\bf 48}, 14099 (1993).

\bibitem{Hoo97}
R. Q. Hood, M. Y. Chou, A. J. Williamson, G. Rajagopal, R. J. Needs,
and W. M. C. Foules,
\prl {\bf 78}, 3350 (1997).

\bibitem{And82}
T. Ando, A. B. Fowler and F. Stern,
\rmp {\bf 54}, 437 (1982).

\bibitem{Ste84}
F. Stern,
\prb {\bf30}, 840 (1984).

\bibitem{Ste74}
F. Stern,
Jpn. J. Appl. Phys. Suppl. {\bf 2}, Pt. 2, 323 (1974).

\bibitem{Kwo93}
Y. Kwon, D. M. Ceperley, and R. M. Martin,
\prb {\bf 48}, 12037 (1993).

\bibitem{Tau93}
M. Taut,
\pra {\bf 48}, 3561 (1993).

\bibitem{Fil94}
C. Filippi, C. J. Umrigar, M. Taut,
\jcp  {\bf 100}, 1290 (1994).

\bibitem{note:EXX}
For the ground state properties of the example, in which only the lowest
single state is occupied, the ``exact exchange'' (EXX) method in DFT is
identical to the Hartree-Fock scheme. Discussion on the physical differences
between the EXX and Hartree-Fock methods and EXX calculation results of
molecular systems are presented in Ref. \onlinecite{YHK4}.

\bibitem{YHK2}
Y.-H. Kim, I.-H. Lee, and R. M. Martin,
 in {\em Stochastic Dynamics and Pattern Formation in Biological and Complex Systems},
 edited by S. Kim, K. Lee, T.K. Lim and W. Sung (AIP, 1999).

\bibitem{YHK4}
Y.-H. Kim, M. St\"adele, and R. M. Martin,
\pra {\bf 40} 3633 (1999).

\bibitem{Kro86}
E. Krotscheck and W. Kohn,
\prl {\bf 57}, 862 (1986).

\bibitem{Aci96}
P. H. Acioli and D. M. Ceperley,
\prb {\bf 54}, 17199 (1996).

\bibitem{Pir98}
J. M. Pitarke and A. G. Eguiluz,
\prb {\bf 57}, 6329 (1998).

\bibitem{Ras75}
M. Rasolt and D. J. W. Geldart,
\prl {\bf 35}, 1234 (1975);
D. J. W. Geldart and M. Rasolt,
\prb {\bf 13}, 1477 (1976).

\bibitem{Ras86}
M. Rasolt and D. J. W. Geldart,
\prb {\bf 34}, 1325 (1986);
\prl {\bf 60}, 1983 (1988).

\bibitem{Lee97}
I.-H. Lee and R. M. Martin,
\prb {\bf 97}, 7197 (1997).

\bibitem{Che77}
R. Chen, P. Trucano, and R. F. Stewart,
Acta. Cryst. A {\bf 33}, 823 (1977).

\bibitem{Pol00}
L. Pollack and J. P. Perdew,
(in preparation).

\end{references}
\end{document}